# A QoS-Aware Intelligent Replica Management Architecture for Content Distribution in Peer-to-Peer Overlay Networks

S.Ayyasamy[1]
[1] Asst. Professor, Department of Information Technology,
Tamilnadu College of Engineering
Coimbatore-641 659, Tamil Nadu, INDIA,
Email: ayyasamyphd@gmail.com

S.N. Sivanandam[2]
[2] Professor and Head, Department of Computer Science
and Engineering, PSG College of Technology,
Peelamedu, Coimbatore-641 004, Tamil Nadu, INDIA.

***Abstract*— The large scale content distribution systems were improved broadly using the replication techniques. The demanded contents can be brought closer to the clients by multiplying the source of information geographically, which in turn reduce both the access latency and the network traffic. The system scalability can be improved by distributing the load across multiple servers which is proposed by replication. If a copy of the requested object (e.g., a web page or an image) is located in its closer proximity then the clients would feel low access latency. Depending on the position of the replicas, the effectiveness of replication tends to a large extent. A QoS based overlay network architecture involving an intelligent replica placement algorithm is proposed in this paper. Its main goal is to improve the network utilization and fault tolerance of the P2P system. In addition to the replica placement, it also has a caching technique, to reduce the search latency. We are able to show that our proposed architecture attains less latency and better throughput with reduced bandwidth usage, through the simulation results.**

## I. INTRODUCTION

### 1.1 Overlay Networks

To share the computer resources like content, storage, CPU cycles directly without using an intermediate system or a centralized server, distributed computer architecture, called "peer-to-peer" are designed. They are distinguished by their failure adaptation capabilities and maintenance of acceptable connectivity and performance [1]. Significant research attention has been applied to Content distribution, which is an important peer-to-peer application on the Internet. By allowing personal computers to work as a distributed storage medium, they normally contribute, search and obtain digital content.

Overlays are flexible and deployable approaches that allow users to perform distributed operations without modifying the underlying physical network. Peer-to-peer (P2P) overlay systems have been proposed to address a variety of problems and enable new applications. The attraction of these systems, when compared to client/server frameworks, is in their robustness, reliability and cost efficiency.

Unlike traditional distributed computing, P2P networks aggregate large number of computers and possibly mobile or handheld devices, which join and leave the network frequently. Nodes in a P2P network, called peers, play a variety of roles in their interaction with other peers. When accessing information, they are clients. When serving information to other peers, they are servers. When forwarding information for other peers, they are routers. This new breed of systems creates application-level virtual networks with their own overlay topology and routing protocols.

To search for data or resources, messages are sent over multiple hops from one peer to another with each peer responding to queries for information it has stored locally. Structured P2P overlays implement a distributed hash table data structure in which every data item can be located within a small number of hops at the expense of keeping some state information locally at the nodes.

### 1.2 Replica Placement for QoS-Aware Content Distribution

Replication techniques are widely employed to improve the availability of data, enhancing performance of query latency and load balancing in content distribution systems. We can geographically multiply the source of information by distributing multiple copies of data in the network. By forwarding each query to its nearest copy, the query search latency can be effectively reduced.

The ability to improve system scalability through distributing the load across multiple servers [2] is also offered by replication. If a replica of the requested object (e.g., a web page or an image) is kept in its nearer proximity then the clients would feel low access latency. Depending on the position of the replicas, the effectiveness of replication tends to a large extent.

The centralized servers become a bottleneck as the requirement of the information increases. The performance problem is managed by the content providers, system administrators or end users by themselves through delivering replicas of web content to machines, spread throughout the network. The load on the central server [3] is reduced by replicas through responding to the local client requests. The load which is delivered to the cooperate nodes includes:

• Communication bandwidth for sending the data to the requesting content

• Storage used for hosting the replica and

• CPU resources for query processing.

The problem of deciding how many replicas is to be delivered to each file and its location is given by the Replica management to this circumstances. To handle more requirements for each file, enough replicas should be present. Servers become overloaded and clients observe lower performance by having only few replicas. On the other hand the waste bandwidth of extra replicas and the storage which could be reassigned to the other files, and also the money spent to rent, power and also for host machine cooling.

In this paper, we propose QOS-aware Intelligent Replica Management (QIRM) architecture for peer-to-peer overlay network. It contains a replica placement algorithm and a robust query searching technique for data retrieval.

This paper is organized as follows. Section 2 gives the detailed related work done. Section 3 presents the system model and algorithm overview for the proposed architecture. Section 4 presents the intelligent replica placement algorithm, followed by the searching technique. Section 5 gives the experimental results and section 6 concludes the paper.

## II. Related Works

Most of the research efforts to improve the performance of Gnutella-like P2P systems can be classified into two categories:

P2P search and routing algorithms and

P2P overlay topologies.

Most of the proposed routing or search algorithms in the first category, disregard the natural peer heterogeneity present in most P2P systems, and more importantly the potential performance hurdle caused by the randomly constructed overlay topology.

B. Mortazavi_ and G. Kesidis [4] have provided a survey of reputation systems. Based on a reputation framework, they have designed a game in which users play to maximize the received files from the system. For this, the users adjust their cooperation level, there by obtaining a good reputation as a result.

Raphael Chand and Pascal Felber have designed for publish or subscribe system based on peer -to - peer paradigm. A containment-based proximity metric was proposed which allows us to build a bandwidth-efficient network topology that produces no false negatives and very few false positives. They have also developed a proximity metric based on subscription similarities which yields a more solid graph structure with negligible false negatives ratios and very few false positives [5].

Anwitaman Datta have discussed some of the important issues concerning structured P2P systems and interplay between the two P2P and MANET self-organizing networks from a data management perspective which aims to achieve efficient and robust information search and access schemes [6].

Paraskevi Raftopoulou and Euripides G.M. Petrakis have presented iCluster, a self-organizing peer-to-peer overlay network for supporting full-fledged information retrieval in a dynamic environment. They defined the criteria for peer similarity and peer selection, and also presented the protocols for organizing the peers into clusters and for searching within the clustered organization of peers [7].

Carvalho, N. Araujo, F. Rodrigues. L, have presented the IndiQoS architecture, a scalable QoS-aware publish-subscribe system with QoS-aware publications and subscriptions that preserves the decoupling which makes the publish-subscribe model so appealing. To support such model, the proposed architecture IndiQoS includes a decentralized message-broker based on a DHT that leverages on underlying network-level QoS reservation mechanisms [10].

Guillaume Pierre and Maarten van Steen have presented Globule, a collaborative content delivery network. The Proposed network was composed of Web servers that cooperate across a wide-area network to provide performance and availability guarantees to the sites they host [12].

Yan Chen et al. [14] have proposed the dissemination tree, a dynamic content distribution system built on top of a peer-to-peer location service. They have presented a replica placement protocol that has built the tree while meeting QoS and server capacity constraints. The number of replicas as well as the delay and bandwidth consumption for update propagation was significantly reduced.

Jian Zhou et al. [15] have shown that the replica placement problem in P2P networks has represented as a Clustered KCenter problem (which essentially differed from the classic kcenter problem) and is proven to be NP-complete. To solve this problem, they bring forward an approximation algorithm in the form of a distance graph for the network topology; when their defined feasibility condition has hold at a certain point; the replica placement solution has built out of (m-1) power of current distance graph.

Kan Hung Wan and Chris Loeser [16] have proposed techniques and algorithms for point-to-point streaming in autonomous systems as it might occur in large companies, a campus or even in large hotels. Their major aim was to create a replica situation that inter-sub network RSVP streams are reduced to a minimum. Therefore, they have introduced the architecture of an overlay network for interconnecting sub networks. Each sub network contains a so-called local active rendezvous server (LARS) which does not just act as directory server but also controls availability of movie content in its subnet work. Due to this, they have considered data placement strategies depending on restrictions of network bandwidth, peer capabilities, as well as the movie's access frequency.

Spiridon Bakiras and Thanasis Loukopoulos [17] have discussed that the caching and replication have emerged as the two primary techniques for reducing the delay experienced by end-users when downloading web pages. They have investigated the potential performance gain by using a CDN server both as a replicator and as a proxy server. They have developed an analytical model to quantify the benefit of each technique, under various system parameters, and they have proposed a greedy algorithm to solve the combined caching and replica placement problem.

Jie Su and Douglas Reeves [18] have proposed that bounding the latency of client requests are an important factor in solving the replica placement problem for content distribution networks. They have proposed two algorithms for placing replicas with latency constraints efficiently, one centralized, and one distributed. They have shown that the impact on the number of replicas required as the latency constraint has become more stringent. In the case where client request patterns were unknown, they have shown that the additional number of replicas needed is reasonable.

Unfortunately, most existing work on replica placement has focused on optimizing an average performance measure of the entire client community such as the mean access latency [8], [9]. While an average performance measure may be important from the system's point of view, it does not differentiate the likely diverse performance requirements of the individuals. So far, to the best of our knowledge, there has been no study on QoS-aware replica placement.

## III. System Model and Algorithm Overview

### 3.1 Algorithm Overview

In our QOS aware topology, nodes are grouped into strong and weak clusters based on their weight vector which comprises the following parameters:

*Available capacity*

*CPU speed*

*Memory size*

*Access Latency*

In the replica placement algorithm, we classify the content as Class I and Class II, based on their access patterns. (i.e.) The most frequently accessed contents are ranked as Class I and the less frequently accessed contents as Class II. Then more copies of Class I content are replicated in strong clusters (having high weight values).

Routing is performed hierarchically by broadcasting the query only to the strong clusters.

Thus our proposed architecture achieves Low bandwidth Consumption, Reduced Latency, Reduced Maintenance Cost, Strong Connectivity and Query Coverage.

### 3.2 System Model

Let us consider a collection of N server nodes which form a peer to peer (P2P) overlay network. In addition to being part of the overlay, each node functions as a server responding to requests (queries) which come from clients outside of the overlay network. An example could be that each node is a web server with the overlay linking the servers and clients being web browsers on remote machines requesting content from the servers.

We assume each node always stores one copy of its own content item which it serves to clients and that it has additional storage space to store k replicated content items from other nodes which it can also serve [3]. The object is associated with an authoritative *origin server* (OS) in the network where the content provider makes the updates to the object. The object copy located at the origin server is called the *origin copy* and an object copy at any remaining server is called a *replica*.

## IV. Intelligent Replica Placement Algorithm

### 4.1 Clustering the Nodes

*For* each node $N_i, i = 1,2.......n$, let

$BW_i$ - Available bandwidth

$SP_i$ - CPU speed

$AL_i$ - Access Latency

$MZ_i$ - Memory Size

1. The weight of the node Ni can be calculated as

$$W_i = (BW_i + SP_i + MZ_i) / AL_i$$

2. Form the vector $W = \{S_i, W_i\}$ , which denotes the node ids and their corresponding weight values, sorted on the descending order.

3. Let {Sk} denote the set of strong cluster nodes $(0 <= k < n)$ , which satisfies the following condition $W_k \geq \beta$, where $\beta$ is the minimum threshold value for the weight.

4. Then the set {Wj} = {Ni} – {Sk}, denote the set of weak cluster nodes $(0 <= j < n)$ , which satisfies the condition $W_k < \beta$

### 4.2 Replica Placement

Let QS be the query server which registers the query of each client. The query server stores the cluster information of each node along with the node id as "S" or "W" for strong and weak clusters, respectively.

At time Tk, let m clients generates query requests {Qm} of the form q{nid, ckwd}, where nid is the node id of the client and ckwd is the keyword of the content to be retrieved.

The queries {Qm} are registered in the query server QS.

The requested content of the queries are classified and categorized as class1 or class2, depending on the access frequencies.

(i.e.) A query Qj, j < m, is considered to be class1

If n (Qj) >= Amin

and class2,

If n (Qj) < Amin

Where n (Qj) is the number of access of the content pattern for the given query and Amin is the minimum access threshold value.

Then the query server QS assigns the class1 contents to the strong cluster nodes and class2 contents to the weak cluster nodes.

After the assignment, QS transmit these replication pattern information to the origin server OS.

OS performs the replication placement, according to the pattern information obtained from QS. The weight value Wi of each node is stored along with the content.

OS then broadcasts the replication information to the respective clients in the following format

{Nid, Clid (“S” or “W”), c1, c2 …}

Where Nid is the node id, Clid is the cluster id and c1, c2… are content database ids.

### 4.3 Query Search and Data Retrieval

A route discovery algorithm is needed to determine if and where the requested item is replicated when the requester does not have knowledge of the destination.

By reducing the communication costs, the speed and efficiency of the information retrieval mechanism can be improved. So, the number of messages exchanged between the nodes and the number of cluster nodes that are queried for each query request, are to be minimized. For this, a robust searching algorithm is proposed.

In this algorithm, each node maintains a profile which contains the details of queries processed by its neighbors, within the last t seconds.

Node Id (Ni)

Query Id (Qid)

Query Hits (Qhit) and

No. of Results (NoR)

This profile information is then used to forward the queries to the neighbors who are having more chances of replying to those queries.

In order to forward a query Q, to its neighbors, a node N1 assigns a score to each of its neighbors based on their profile. To calculate the score of each node Nj, (j=2, 3…) N1 compares .Q with all queries stored in Nj’s profile. If there is a query hit for Q, then the score of Nj can be calculated as

$$\text{Score}(Nj,Q) = \sum_{k=0}^{m} NoR(Nj,Qk)^{\alpha}$$

Where NoR (Nj, Qk) is the number of results returned by Nj for query Qk, which are similar to Q. So the nodes which return more results get the higher score.

If α allows us to add more weight to the most similar queries. For example, when α is large, then the query with the largest similarity NoR (Nj, Qk) dominates the formula.

If we set α = 1, all queries are equally counted, whereas setting α=0 allows us to count only the number of results returned by each peer.

(i) When a data request is initiated at a client, it first looks for the data item in its own cache (local hit). If there is a local cache miss, the client sends the *request to* the set of strong cluster nodes.

(ii) On receiving the request, each strong cluster node which has the requested content, will send an ack packet to the query client to acknowledge that it has the data item. The ack packet will contain the following fields: time stamp Ts and weight W. The time stamp field helps to choose the latest copy of the searched item and the weight value field helps to choose the best client node.

(iii) When the query client receive *ack* packets from the strong cluster, it selects the best node Sbest with max $(T_s, W)$ and sends a confirm packet to the client Sbest. The *ack* packets for the same item received from other nodes are discarded

(iv) When the node Sbest receives a *confirm* packet, it responds back with the actual data value to the requested query node.

(v) Suppose if the requested data is not available in any strong cluster nodes, the request is directed to the server from the query client. Then the necessary data is sent to the client from the server. If the client has the available memory size (MZ) and bandwidth (BW), then it caches the data in its buffer. Then it is also considered as a strong cluster node and it is propagated to other nodes as

{Nid, Clid (“S”), d1}

Where Nid is the node id, Clid is the cluster id and d1… is the content database id.

(vi) Subsequently if the same data is required for any other client, then it sends its request to this strong cluster node which caches the data and receives the required data.

Caching frequent data which is not found in the replica, into the local cache of a node increases the query efficiency and decreases the latency significantly.

We now summarize the above steps into the following algorithm.

**Algorithm**

1. The Node N1 gets a query Q from a client C.
2. N1 compares Q with $Qid_{kj}$, where $Qid_{kj}$ is the $k^{th}$ query of node Nj.
   k=1, 2, 3…. and j=1, 2…..
3. If $Qhit_k > 0$ , then
4. $\text{Score}(Nj,Q) = \sum NoR(Nj,Qk)^{\alpha}$

   i. k=0

5. Select the Nodes Nj where Score (Nj, Q) is maximum.
6. Forward the query Q to Nj.
7. Nj sends ACK to client C.
8. C selects the node Sbest from Nj, with $\max(T_s, W)$.
9. C sends a confirm packet to Sbest.
10. Sbest send requested data for query Q, to C
11. If $Qhit_k > 0$ for all j, then
12. C send query request Q to Server.
13. Server sends requested data for query Q, to C
14. If $MZ_C > \min(MZ)$ and $BW_C > \min(BW)$ (Where $MZ_C$ – Memory size of C and $BW_C$ –
    i. Bandwidth of C) then
15. C caches the data item.
16. C becomes a strong cluster node.
17. C propagates {Nid , Clid (“S”) , d1 } to other nodes
    (Where Nid is the node id, Clid is the cluster id and d1 is the content database id).

## V. Experimental Results

### 5.1 Simulation Setup

This section deals with the experimental performance evaluation of our algorithms through simulations. In order to test our protocol, the NS2 simulator is used. NS2 is a general-purpose simulation tool that provides discrete event simulation of user defined networks.

We have used the Bittorrent packet-level simulator for P2P networks [13]. A network topology is only used for the packet-level simulator. Based on the assumption that the bottleneck of the network is at the access links of the users and not at the routers, we use a simplified topology in our simulations.

We model the network with the help of access and overlay links. Each peer is connected with an asymmetric link to its access router. All access routers are connected directly to each other modeling only an overlay link. This enables us to simulate different upload and download capacities as well as different end-to-end (e2e) delays between different peers.

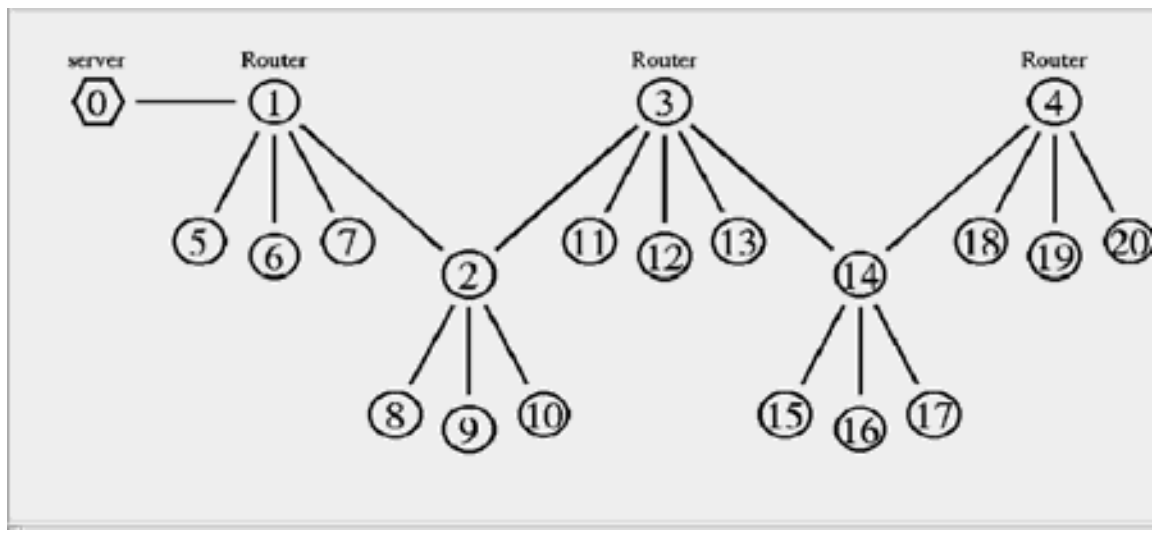

Figure 1. Topology of P2P Overlay Network

### 5.2 Simulation Results

We have compared our QIRM architecture to Virat, a node capability aware P2P middleware [11] architecture for managing replicas in large scale distributed systems.

**Based On Load**

In our initial experiment, the load of the requested content is varied from 2.0mb to 5.0 Mb. The response delay in seconds and received throughput in packets are measured. In Figure 2, we can see that, when the load increases, the delay also increases. It is evident that the delay of QIRM is significantly less than the delay of VIRAT.

Figure 3 shows the aggregated throughput of all the client nodes which obtained their respective share of data. From the figure we can see that the QIRM has more throughput than VIRAT and the throughput values are decreasing, when the load increases.

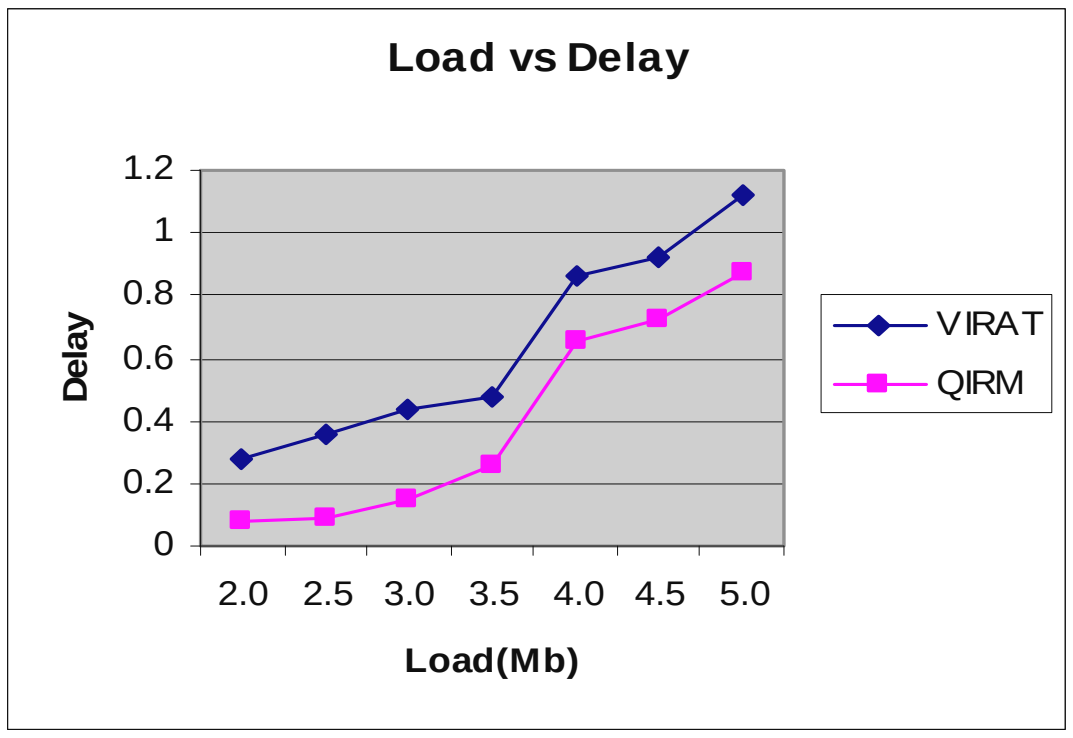

Figure 2. Load Vs Delay (s)

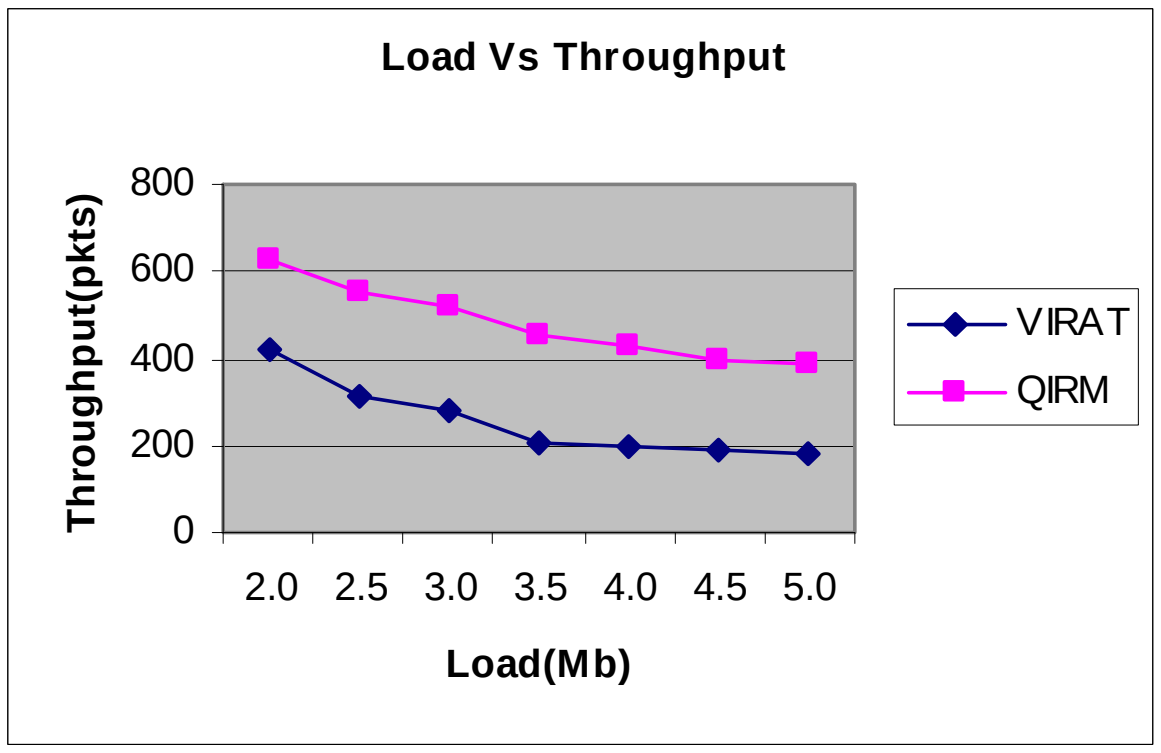

Figure 3. Load Vs Throughput

**Based On Rate**

In our second experiment, the query sending rate is varied from 250Kb to 1Mb. The response delay in seconds and query efficiency are measured. Query efficiency is a measure of the percentage of data queries that get served during an entire simulation.

Figure 4 shows that the average query efficiency of the client nodes increases when the rate is increased. From the figure, we can see that the QIRM has more efficiency than VIRAT.

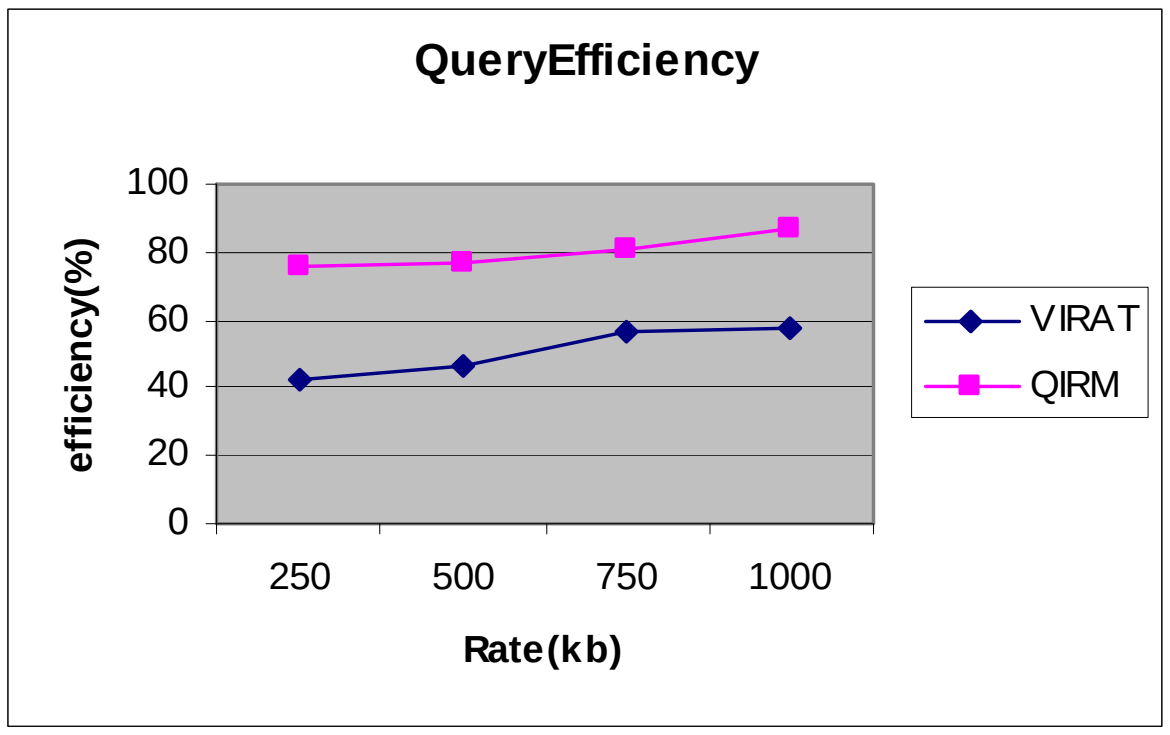

Figure 4. Rate Vs Packet Delivery Fraction

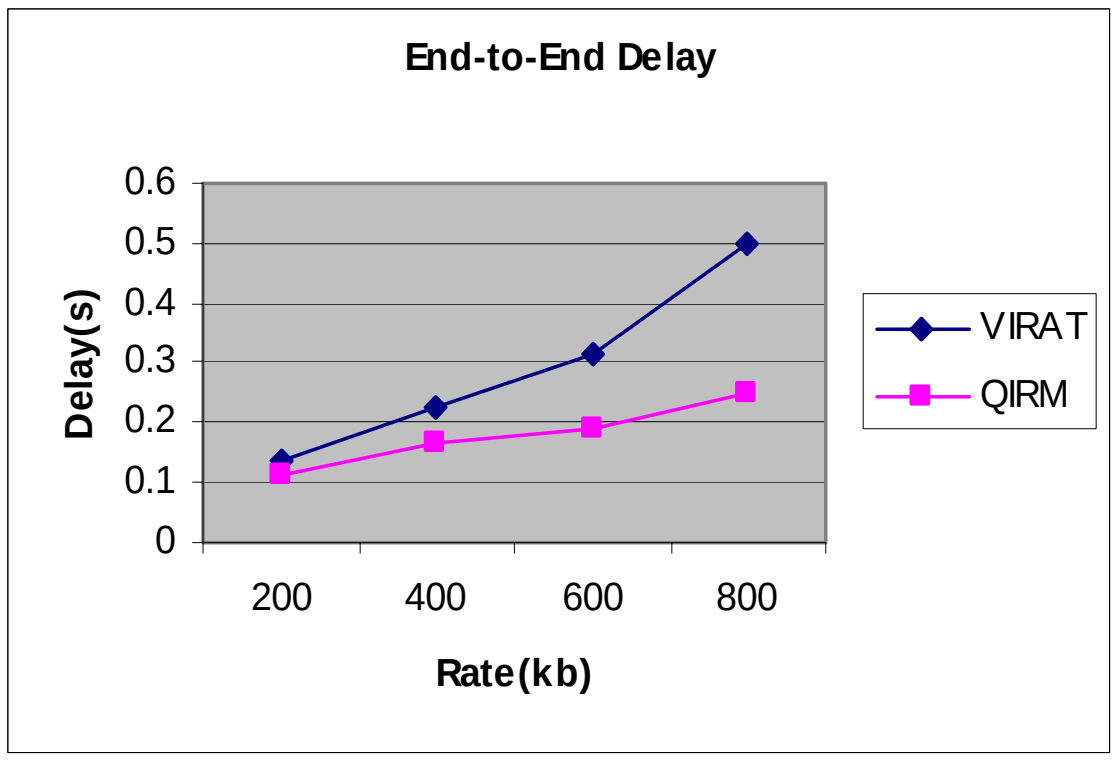

Figure 5. Rate Vs Delay

In Figure 5, we can observe that, when the rate increases, the delay remains almost constant. From the figure, it can be seen that the delay of QIRM is significantly less than the delay of VIRAT.

In Figure 6, the throughput against rate is shown. From the figure, we can see that the throughput of QIRM is more when compared to VIRAT, and increases when rate increases.

In Figure 7, the bandwidth utilization of clients against the rate is shown. From the figure, we can see that, bandwidth utilization of QIRM is nearly 80-90%, when compared to VIRAT, which is 60-70%.

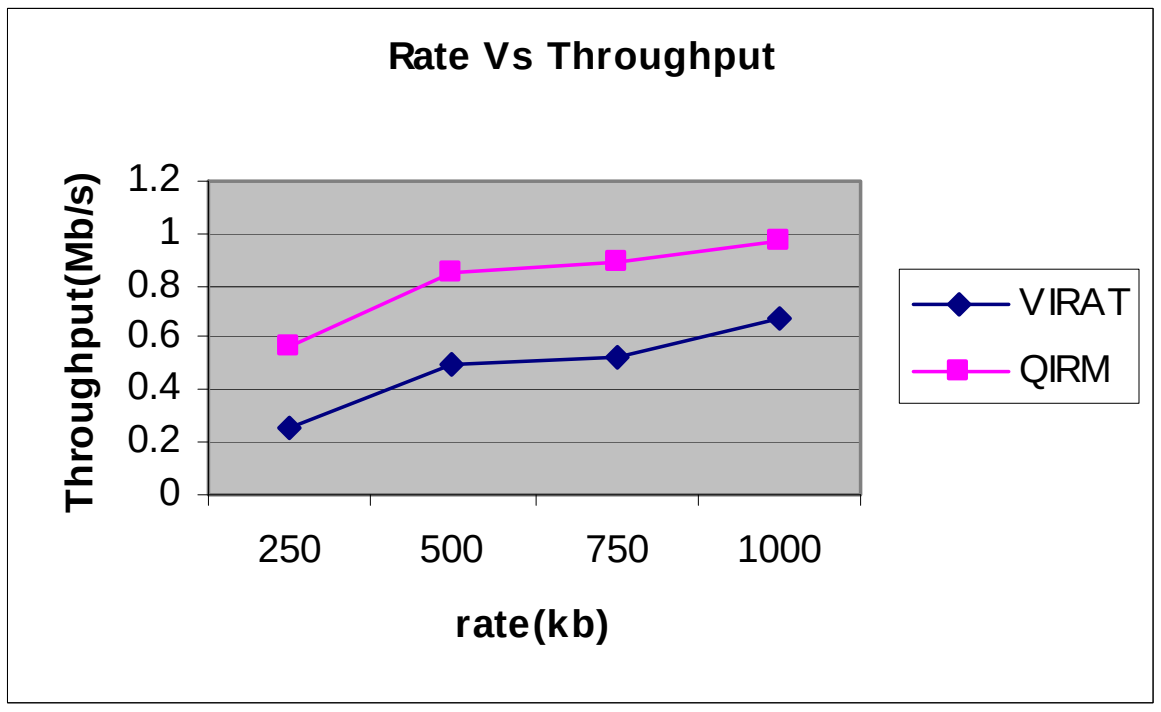

Figure 6. Rate Vs Throughput

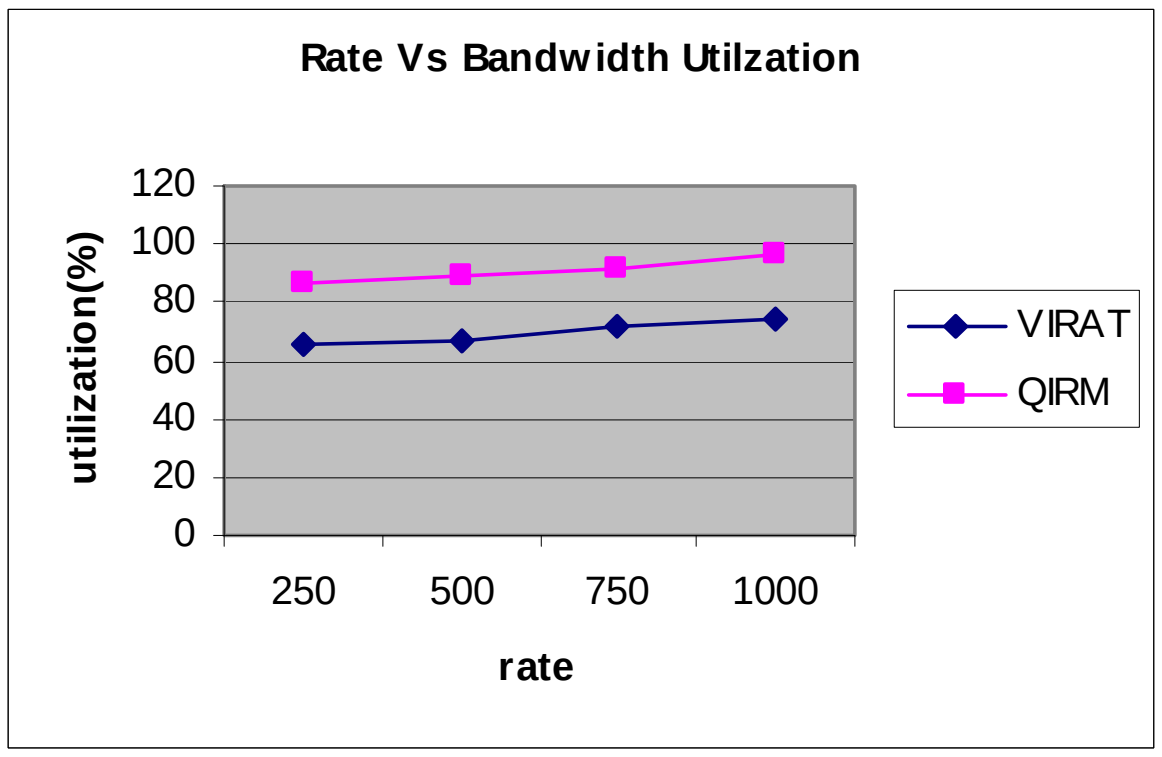

Figure 7. Rate Vs Utilization

## VI. Conclusion

A QoS based overlay network architecture including an intelligent replica placement algorithm is used to improve the network utilization and the fault tolerance of the P2P and also to reduce the search latency. Based on the weight vector which includes available capacity, CPU speed, and memory size and access latency the nodes are classified into strong and weak clusters. Based on the access pattern the content is classified into class I or class II by the replica management algorithm. Then class I contents are replicated into strong groups for more copies. Routing is performed only to the strong clusters through broadcasting the query hierarchically. In addition to the replica placement, it also has a caching technique, to reduce the search latency. Low bandwidth Consumption, Reduced Latency, Reduced Maintenance Cost, Strong Connectivity and Query Coverage are achieved in the proposed architecture. Thus we have shown that our proposed architecture attains less latency and better throughput with reduced network bandwidth usage, through the simulation results.

## About the Authors

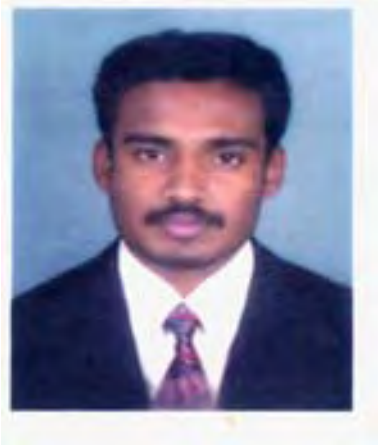

Mr.S.Ayyasamy completed his B.E. (Electronics and Communication Engineering) in 1999 from Maharaja Engineering College and M.E. (Computer Science and Engineering) in 2002 from PSG College of Technology, both under Bharathiar University, Coimbatore. Currently he is pursuing PhD degree from Anna University, Coimbatore. He is working as a Assistant Professor, Department of Information Technology at Tamilnadu College of Engineering, Coimbatore. He is a member of CSI. His research areas include P2P networks, Overlay Networks, Cloud computing and Quality of Services and having 8 years of teaching experience in Engineering Colleges.

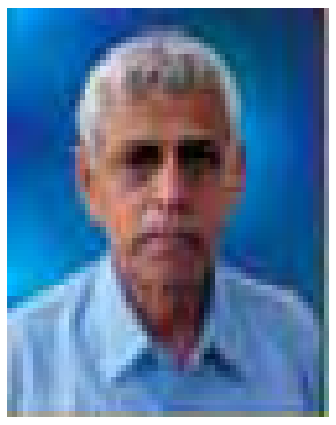

Dr. S. N. Sivanandam completed his B.E. (Electrical Engineering) in 1964 from Government College of Technology, Coimbatore, and MSc (Engineering) in Power Systems in the year 1966 from PSG College of Technology, Coimbatore. He acquired PhD in control systems in 1982 from Madras University. He received best teacher award in the year 2001 and Dhakshina Murthy Award for teaching excellence from PSG College of technology. He received the citation for best teaching and technical contribution in the year 2002, Government College of Technology, Coimbatore. His research areas include Modeling and Simulation, Neural Networks, Fuzzy Systems and Genetic Algorithm, Pattern Recognition, Multidimensional system analysis, Linear and Non linear control system, Signal and Image processing, Control System, Power System, Numerical methods, Parallel Computing, Data Mining and Database Security. He is a member of various professional bodies like IE (India), ISTE, CSI, ACS and SSI. He is a technical advisor for various reputed industries and engineering institutions.